\documentclass[twocolumn,pre,superscriptaddress,showpacs,byrevtex]{revtex4}
\usepackage{epsf,graphicx,amssymb}

\begin{document}
\title{Fluctuations of energy flux in a simple dissipative out-of-equilibrium system}
\author{Claudio Falc\'on}
\affiliation{Laboratoire de Physique Statistique, \'Ecole Normale Sup\'erieure, CNRS, 24, rue Lhomond, 75 005 Paris, France}
\author{Eric Falcon}
\affiliation{Laboratoire Mati\`ere et Syst\`emes Complexes (MSC), Universit\'e Paris Diderot, CNRS (UMR 7057)\\10 rue A. Domon \& L. Duquet, 75 013 Paris, France}
\date{\today}

\begin{abstract} 
We report the statistical properties of the fluctuations of the energy flux in an electronic RC circuit driven with a stochastic voltage. The fluctuations of the power injected in the circuit are measured as a function of the damping rate and the forcing parameters. We show that its distribution exhibits a cusp close to zero and two asymmetric exponential tails, the asymmetry being driven by the mean dissipation. This simple experiment allows to capture the qualitative features of the energy flux distribution observed in more complex dissipative systems. We also show that the large fluctuations of injected power averaged on a time lag do not verify the Fluctuation Theorem even for long averaging time. This is in contrast with the findings of previous experiments due to their small range of explored fluctuation amplitude. The injected power in a system of $N$ components either correlated or not is also studied to mimic systems with large number of particles, such as in a dilute granular gas.
\end{abstract}
\pacs{05.40.-a, 05.70.Ln, 05.10.Gg, 84.30.-r}

\maketitle

 \section{Introduction}
 \label{intro}

Global quantities (i.e., quantities averaged over the boundaries or the entire volume of the system) are of paramount importance to describe the dynamics and the statistics of dissipative out-of-equilibrium systems \cite{Aumaitre99}. For instance, the injected power is necessary to maintain a dissipative system in an out-of-equilibrium regime. A system thus reaches a stationary non-equilibrium state when a balance between the mean injected power and the mean dissipated power is achieved. The usual tools of equilibrium statistical mechanics do not apply to such systems or are reduced to the Fluctuation--Dissipation theorem \cite{VanKampen}. Even more, the injected power $I$ is a fluctuating quantity and cannot be only regarded as a constant parameter. Their fluctuations display values that can be several times larger than its average, and their statistics (even averaged over a macroscopic volume) present large deviations \cite{Aumaitre99,Fauve,Aumaitre04}. It is thus of crucial interest to study the statistical properties of the energy flux driving a system far from equilibrium, and its relation to its internal energy. In some systems, the energy flux fluctuations can be directly related to the internal energy by means of the Fluctuation Theorem (FT).

The FT is of fundamental importance for microscopic systems far from equilibrium in a stationary state. It was first introduced numerically for a fluid under an external shear \cite{Evans93}, then mathematical proof was given \cite{Evans94,GCohen}. For a nonequilibrium dissipative system, this theorem describes the asymmetry of distribution of a fluctuating global quantity $I_\tau$ (energy flux, entropy production rate,...) averaged over a time $\tau$ much larger than its typical correlation time $\tau_c$. For systems close to equilibrium or for macroscopic ones, the FT gives a generalization of the second law of thermodynamics, and also implies the Green-Kubo relations for linear transport coefficients when combined with the central limit theorem \cite{Searles00}. Moreover, it can be applied to nonequilibrium transitions between two different equilibrium states leading to the so-called Jarzynski equality \cite{Reid05}. Its derivation requires the assumption of time reversibility of the system dynamics, ergodic consistency, and a certain initial distribution of particle states. Finally, it does not require or imply that the distribution of time averaged fluctuating quantity $I_\tau$ is Gaussian.

Experimental tests of the fluctuation theorem relation have been reported in various systems: in granular gases \cite{Feitosa04}, in turbulent flows (thermal convection \cite{Ciliberto98,Shang05}, swirling flows \cite{Ciliberto04}), in liquid crystals \cite{Goldburg01}, with an electric dipole \cite{Garnier04} or a mechanical oscillator \cite{Douarche05}, in a two-level atomic system \cite{Schuler05}, and by means of a colloid particle \cite{Wang02} or an RNA molecule \cite{Collin05} in an optical trap. In all these experiments, the fluctuation theorem is found to be verified with good accuracy despite some of these systems do not satisfy the microscopic reversibility hypothesis. Such a good agreement has been also reported in numerical simulations of granular gases \cite{Aumaitre99,Aumaitre04}, turbulence \cite{Aumaitre99,Gilbert04}, and earthquakes \cite{Aumaitre99}. The reasons of this apparent verification of the FT are two fold: either due to the small range of explored fluctuation amplitude $\epsilon \equiv I_\tau/\langle I\rangle$  \cite{Aumaitre99,Aumaitre04,Gilbert04}, or due to the long averaging time $\tau$ needed \cite{Aumaitre99,Aumaitre04,Puglisi05}. Only small relative fluctuation amplitudes ($\epsilon \leq 0.8$ for $\tau \leq 20\tau_c$) have been reached in the above experiments  \cite{Feitosa04,Ciliberto98,Shang05,Ciliberto04}. Very recently, large range of $\epsilon$ has been attained, even for $\tau>>\tau_c$, by measuring the fluctuating injected power in an experiment of wave turbulence on a fluid surface \cite{WT}. This experiment then shows that the FT is not satisfied for high enough $\epsilon$. Such a disagreement was also predicted theoretically in a system described by a Langevin equation \cite{Farago02}. Note that the breakdown of FT has been recently reported numerically \cite{Gilbert04} or theoretically \cite{breaking} in other systems.

In this paper, the fluctuations of energy flux in an electronic circuit are measured to test the fluctuation theorem within a large range of accessible value of fluctuation amplitude ($\epsilon \simeq 3$) even for long averaging time ($\tau/\tau_c\simeq 20$). The electronic circuit is a resistor of resistance $R$ in series with a capacitor of capacitance $C$ driven with a stochastic voltage. This circuit can be view as an electronic analogue of the Langevin equation \cite{Labbe} which describes usually the brownian motion of a particle \cite{langevin08}. It is important to notice that in our experiment the dissipation is selected by the system itself. No {\it ad-hoc} dissipation or thermostat is introduced to ensure the FT hypothesis (i.e. the time reversibility of the system). The study of the statistical properties of the injected power in such a circuit point out three important results:

First, the probability density function (PDF) of the fluctuations of the injected power in the circuit is studied as a function of the control parameters (damping rate, amplitude of the stochastic forcing). The asymmetry is driven by the damping rate: The more the mean dissipation increases, the less the negative events of injected power occur. This electronic circuit is one of the simplest system to understand the properties of the energy flux fluctuations shared by other dissipative out-of-equilibrium systems (such as in granular gases \cite{Feitosa04}, wave turbulence \cite{WT} and convection \cite{Shang05,Shang03}).

Second, we show that the fluctuations of injected power averaged on a time $\tau$ do not verify the fluctuation theorem at large values of $\epsilon$, even for $\tau>>\tau_c$. This occurs for values of $\epsilon$ larger than the most probable value of the injected power PDF. This electronic circuit thus appears to be a very useful tool to test fluctuation theorem in the different limits of the averaging time and of the fluctuation amplitude.

Third, the injected power in a system constituted by an ensemble of $N$ uncorrelated components is then studied. This mimics a dissipative multi-component system driven out-of-equilibrium without spatial correlation between them. The fluctuations of the time averaged injected power of the $N$ components then verifies the fluctuation theorem for finite time. This bridges the gap between results about the test of the FT for systems with low particle number (such as the ones described by the Langevin equation), and systems with large number of uncorrelated particles (such as in a dilute granular gas). The link between them can be understood as a consequence of the central limit theorem.

It is well known that electronic circuits are very useful analogue experiments to study stochastic nonlinear problems \cite{Luchinsky98}. However, one could wonder their relevance with respect to numerical simulations. Analogue circuits get the advantages that any naturally occurring noise necessarily has a finite correlation time, and thus avoid  to pre-select a correlated noise type (Ito-Stratanovivch dilemma) in writing the numerical code \cite{Luchinsky98}. Moreover, the simulation leads to the accumulation of truncation errors, and it takes a longer time to implement and to compute \cite{Luchinsky98}.

The paper is organized as follows. Section \ref{sec:setup} explains the experimental setup of the $RC$ circuit. Section \ref{sec:results} contains the results about the statistical properties of the injected power in the circuit.  Some of the experimental results of the Section \ref{sec:results} are then recovered in Section \ref{sec:model} with a simple model based on a Langevin equation with a Gaussian colored noise (the so-called Orstein-Ulhembeck noise) \cite{WT,Aumaitre08}. Section \ref{sec:FT} contains the experimental test of the Fluctuation Theorem for the energy flux in a $RC$ circuit. Finally, Section \ref{sec:Nsyst} is devoted to the experimental study of the injected power in a system constituted by a set of $N$ uncorrelated components, as well as the test of the Fluctuation Theorem for its energy flux.

 \begin{figure}[h]
 \centering
 \includegraphics[width=0.3\textwidth]{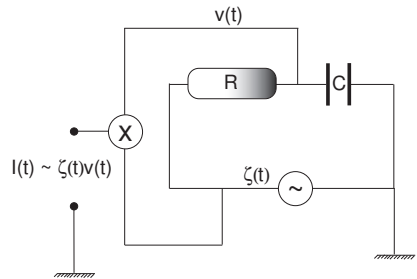}
 \caption{Scheme of the electronic circuit as an analogue of the Langevin equation.}
 \label{Schema}
\end{figure}

\section{Experimental setup}
\label{sec:setup}
The experimental setup consists of a resistor, $R$, in series with a capacitor, $C$, driven with an external stochastic voltage $\zeta(t)$ as displayed in Fig.\ \ref{Schema}. The equation of continuity for the voltage $V$ across the resistor $R$ leads to
\begin{equation}
      \gamma^{-1}\frac{dV(t)}{dt}+V(t) = \zeta(t),
\label{analogueLang}
\end{equation}
where $\gamma^{-1}=RC$. As it will be shown below, the injected power in the $RC$ circuit is
\begin{equation}
I(t) \equiv \gamma V(t) \zeta(t).
\label{power}
\end{equation}
The zero mean Gaussian random noise $\zeta(t)$ is generated by a Spectrum Analyzer (Hewlett-Packard HP 35670A). This noise is low-pass filtered at a cut-off frequency $\lambda$ fixed to 5 kHz, unless specified otherwise. The control parameter is the noise amplitude $D$ defined by the constant value of its power spectral density, as an analogy to the white noise limit. $C$ is fixed to 1 $\mu F$, and $R$ can be varied between 200 $\Omega$ and 10 k$\Omega$ leading to values of $\gamma$ between 50 Hz and 10 kHz. The output $V(t)$ of the $RC$ circuit is multiplied by the random forcing $\zeta(t)$ by means of an analog multiplier (Analog Devices AD540). The resulting voltage $V(t)\zeta(t)$ is proportional to the injected power (see below) and is acquired with a Digital-to-Analog Acquisition card (AT-MIO-16X) at 100 kHz sampling frequency for 10 s, with a precision of 0.3 mV.

Equation\ (\ref{analogueLang}) is the analogue of the Langevin equation that usually describes the dynamics of a Brownian particle of velocity $v$ as \cite{langevin08}
\begin{equation}
\frac{dv(t)}{dt}+\tilde{\gamma} v(t) = f(t)  ,
\label{langevin}
\end{equation}
where $\tilde{\gamma}$ is the inverse of a damping time. $f$ is an external Gaussian random forcing with zero mean and a given autocorrelation function. In the singular limit of zero-correlation time (i.e., for a white noise forcing), this function reads $<f(t)f(t^{\prime})>=f_0\delta(t-t^{\prime})$ and the Fluctuation--Dissipation theorem is satisfied with $\left<v^{2}\right>=f_0/(2\gamma)$, $f_0$ being the noise intensity \cite{VanKampen}. For a non-zero correlation time (as in this study), the system cannot be in equilibrium, and another viscous term different from the one of Eq.\ (\ref{langevin}) must be used to recover the equilibrium state \cite{Jean2}. Multiplying Eq.\ (\ref{langevin}) by $v$ gives
\begin{equation}
\frac{d}{dt}\left[\frac{v(t)^{2}}{2}\right]= f(t) v(t) - \tilde{\gamma} v(t)^{2}{\rm \ ,}
\end{equation}
meaning that the energy budget of the system is driven by the injected power, $f(t) v(t)$, and the dissipative one, $\tilde{\gamma} v(t)^{2}$. This analogy thus shows easily that Eq.\ (\ref{power}) is the injected power in the electronic circuit.

The aim is now to study the probability distribution function (PDF) of the injected power in the $RC$ circuit, described by a Langevin equation as the simplest dissipative system driven out of equilibrium by an external force. The objective is to probe the out-of-equilibrium statistical properties of the injected power and its relation with the fluctuation theorem. It is noteworthy to underline that in this simple system the forcing $f(t)$ is not in any case a thermal bath. Due to the non-zero correlation time of the forcing, this system is strongly out of equilibrium and the Fluctuation--Dissipation theorem does not hold \cite{VanKampen}. This is mainly due to the non-Gaussian shape of the injected power distribution, in contrast with other experimental devices where the injected power fluctuations are quasi normal \cite{Garnier04,Douarche05}.

\section{Statistical properties of the injected power}
\label{sec:results}
\begin{figure}[h]
\centerline{
  \epsfysize=70mm
  \epsffile{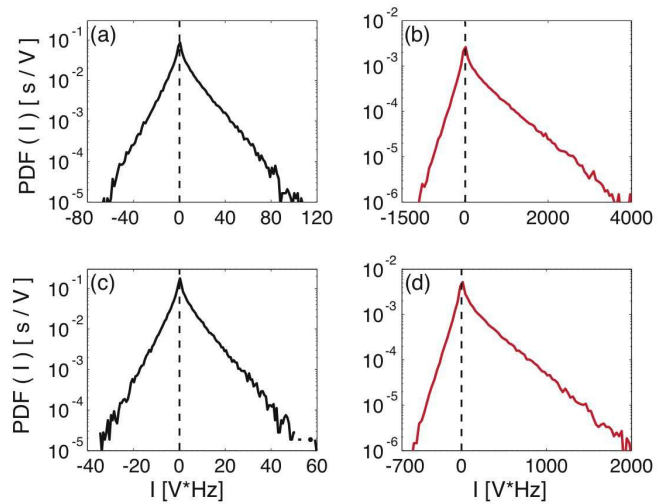}
}
\caption{(Color online) Probability density functions of the injected power $I$ for two different noise amplitudes: $D=1.56 \times 10^{-3}$ V$^2_{\rm rms}$/Hz [(a) and (b)], $D=0.75\times 10^{-3}$ V$^2_{\rm rms}$/Hz [(c) and (d)]; and damping rates: $\gamma= 200$ Hz [(a) and (c)],  $\gamma=2000$ Hz [(b) and (d)].}
\label{PDFS}
\end{figure}

The probability density function of the injected power, $I(t)$, is shown in Fig.\ \ref{PDFS} for different values of the noise amplitude $D$, and the damping rate $\gamma$. For all values of $D$ and $\gamma$, the PDFs exhibit two asymmetric exponential tails and a cusp near $I\simeq 0$. Note that this typical PDF shape has been also observed in various more complex systems (granular gases \cite{Feitosa04}, wave turbulence \cite{WT} and convection \cite{Shang05,Shang03}). As shown in Fig.\ \ref{PDFS}, the PDF skewness increases strongly with $\gamma$ at a fixed $D$. Moreover, the extremal fluctuations increase strongly with $D$ at a fixed $\gamma$.
\begin{figure}[h]
\centerline{
  \epsfysize=70mm
  \epsffile{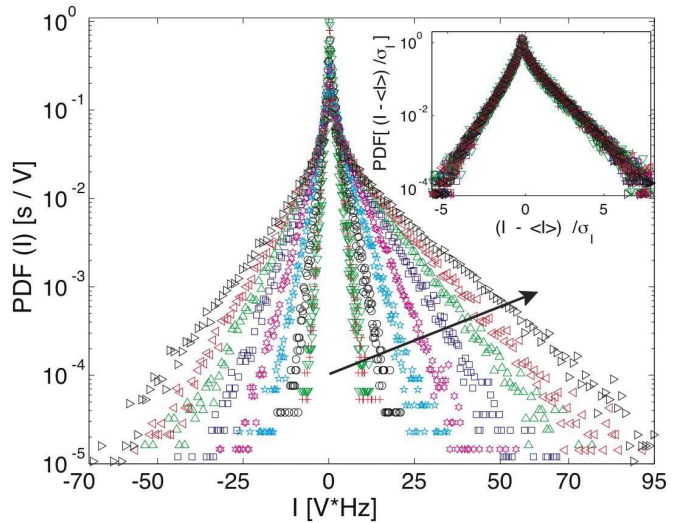}
}
 \caption{(Color online) Probability density functions of injected power, $I$, for $D=0.06$ to $1.56 \times 10^{-3}$ V$^2_{\rm rms}$/Hz (see the arrow) for $\gamma=200$ Hz.  Inset:  Probability density functions in the rescaled variable $(I - \langle I \rangle)/\sigma_I$.}
 \label{PDFcr}
\end{figure}

\begin{figure}[h]
\centerline{
  \epsfysize=70mm
  \epsffile{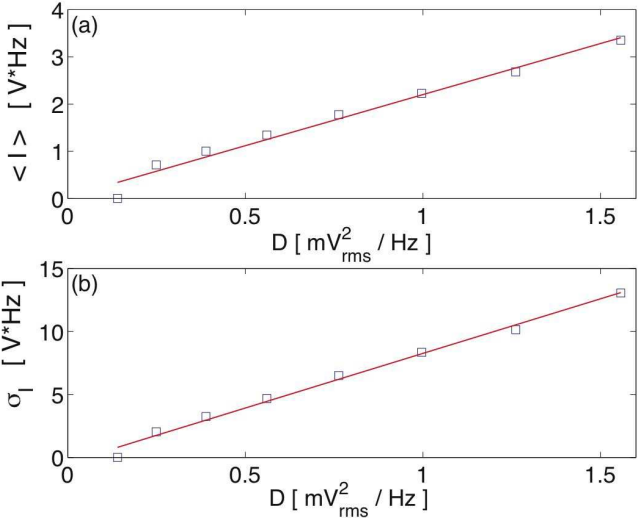}
}
\caption{(Color online) Mean $\langle I \rangle$ and standard deviation $\sigma_{I}$ of the injected power as a function of the noise amplitude $D$. $\gamma=200$ Hz.}
 \label{ScalingD}
\end{figure}

At a fixed value of $\gamma$, the PDFs of $I$ are plotted in Fig.\ \ref{PDFcr} for 9 different noise amplitudes. As shown in the inset of Fig.\ \ref{PDFcr}, all these PDFs collapse on the same curve when plotted in the centered-reduced variable, $(I - \langle I \rangle)/\sigma_I$, where $\sigma_I$ is the rms value of $I$, and $\langle I \rangle$ its mean value. Such a collapse means that all the moments of $I$ scale as $\sigma_I$.  As shown in Fig.\ \ref{ScalingD}, $\sigma_I$ (as well as $\langle I \rangle$) scales linearly with $D$. This linear dependence with $D$ of the PDF of $I$  can be recovered by dimensional analysis. Thus, since the slopes of the exponential tails scale as $D^{-1}$, when the noise amplitude $D$ is doubled, the largest injected power fluctuation reached is doubled.

\begin{figure}[h]
\centerline{
  \epsfysize=70mm
  \epsffile{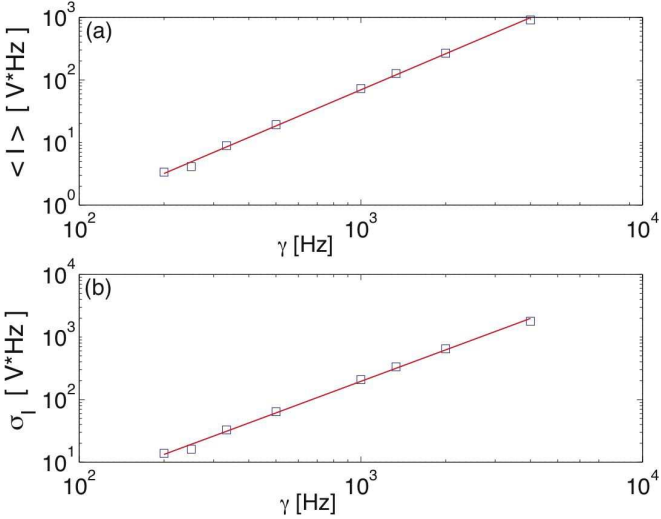}
}
\caption{(Color online) Mean $\langle I \rangle$ and standard deviation $\sigma_{I}$ of the injected power as a function of the damping rate $\gamma$. $D=0.75 \times 10^{-3}$ V$^2_{\rm rms}$/Hz. ($-$): linear best fits of slopes 1.9 V and 1.59 V, respectively.}
 \label{ScalingGamma}
\end{figure}

\begin{figure}[h]
\centerline{
  \epsfysize=70mm
  \epsffile{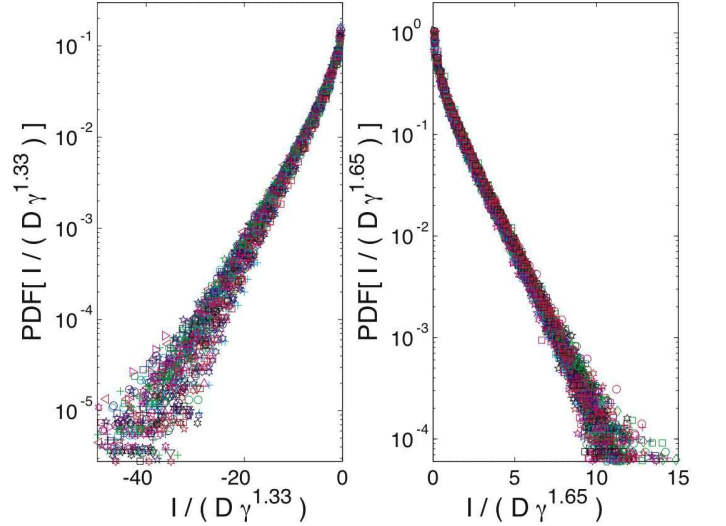}
}
  \caption{(Color online) Scaling of the PDFs of the negative values (left) and the positive values (right) of injected power $I$, for 9 values of $D$, and 10 values of $\gamma$.}
  \label{PDF+-}
  \end{figure}

The noise amplitude $D$ is now fixed in order to take into account the effect of the damping rate $\gamma$ on the injected power fluctuations.  For different values of $\gamma$, $\langle I \rangle$ and $\sigma_I$ are plotted in Fig.\ \ref{ScalingGamma}.  Both moments scale as a power law of $\gamma$ with two different exponents. Therefore no collapse occurs when the PDFs of $I$ are plotted in the centered-reduced variable. However, as displayed in Fig.\ \ref{PDF+-}, both the exponential tails of positive and negative values of $I$ show power law dependences with $\gamma$. The slope of the positive exponential tails scales as $\sim\gamma^{-1.65 \pm 0.05}$, whereas the negative one scales as  $\sim\gamma^{-1.33  \pm 0.05}$. This means that the probability of having negative values of injected power decreases faster than the probability of having positive ones as the system becomes more and more dissipative. 

Taking into account both the effect of $D$ and $\gamma$, the PDF of the positive values of $I$ behaves, far from the cusp at $I\simeq0$, as
\begin{equation}
        P_{+}(I)\sim \exp{\left(-\alpha_{+} \frac{I}{D\gamma^{1.65}}\right)}.
\end{equation}
Similarly, the PDF of the negative values of $I$ behaves as
\begin{equation}
        P_{-}(I)\sim \exp{\left(\alpha_{-} \frac{I}{D\gamma^{1.33}}\right)}
\end{equation}
where $\alpha_{\pm}$ are two constants. As shown below in Sect.\ \ref{sec:model}, an explicit formula of the PDF of $I$ can be computed \cite{WT}, that can capture the properties of the distribution found here: a cusp close to zero and asymmetric exponential tails (see Sect.\ \ref{sec:model}).

\begin{figure}[h]
\centerline{
  \epsfysize=70mm
  \epsffile{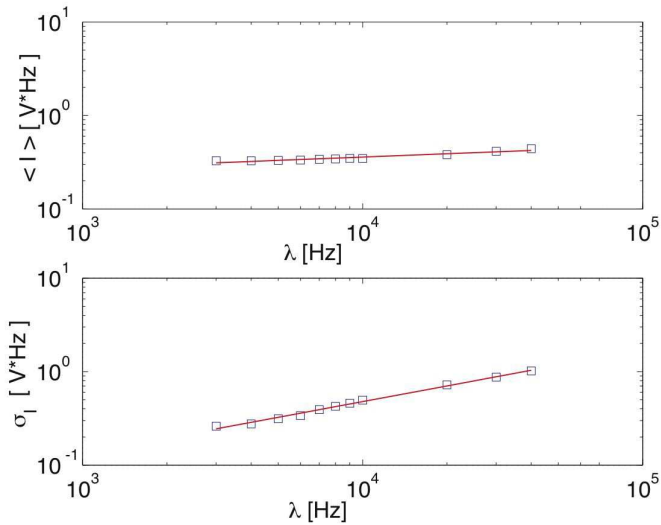}
}
 \caption{(Color online) Scaling of the mean $\langle I \rangle$ and standard deviation $\sigma_{I}$ with the cut-off frequency $\lambda$. ($-$): linear best fit of slope 0.11 V and 0.56 V, respectively.}
 \label{lambda}
 \end{figure}

Both $D$ and $\gamma$ are now fixed in order to study the effect of the random noise cut-off frequency $\lambda$ on $\langle I \rangle$ and $\sigma_{I}$.  As shown in Fig.\ \ref{lambda}, when $\lambda$ is varied from 3 kHz to 40 kHz, the mean injected power is roughly found independent of $\lambda$ with our experimental accuracy, whereas $\sigma_{I}$ scales as the square root of $\lambda$.

Finally, to summarize all the experimental results, the two first moments of injected power behave like
\begin{eqnarray}
\langle I \rangle \sim D \gamma^{1.90} \ \ {\rm and}\ \
\sigma_{I} \sim D\gamma^{1.59}\lambda^{0.50}.
\label{expresult2moments}
\end{eqnarray}
Note that all the previous exponents are experimentally measured with an accuracy of $\pm 0.05$. Thus, the noise amplitude $D$ is found to drive the scale of the injected power fluctuations whereas the damping rate $\gamma$ controls the asymmetry of the PDF of $I$.
 
\section{Langevin-type model with an Orstein-Ulhembeck forcing}
\label{sec:model}
Using a simple model that has been recently presented in Ref.\ \cite{WT} and will be discussed in details in another paper \cite{Aumaitre08}, let us try to recover the above experimental results: The shape of the injected power distribution, and the scaling of its first cumulants ($\langle I \rangle$ and $\sigma_{I}$) with the parameters $D$, $\gamma$ and $\lambda$. 

From Eq.\ (\ref{analogueLang}) of the electronic circuit, and the fact that the stochastic forcing $\zeta(t)$ is low-pass filtered at frequency $\lambda$, one can write the following coupled linear equations
\begin{eqnarray}
\frac{dV(t)}{dt}+\gamma V(t) = \gamma\zeta(t) \label{model1} \\
\frac{d\zeta(t)}{dt}+\lambda \zeta(t) = \xi(t)
\label{model2}
\end{eqnarray}
with $V(t)$ the voltage, $\gamma^{-1}=RC$ the damping parameter, $\zeta(t)$ the colored random forcing, and $\xi(t)$ the Gaussian white noise with $<\xi(t)\xi(t^{\prime})>=D\delta(t-t^{\prime})$, $D$ being the noise amplitude. Note that if we only used a Gaussian white noise in Eq.\ (\ref{model1}), then one find $\langle I \rangle \sim \sigma_{I} \sim D$ but with no dependence with $\gamma$ \cite{VanKampen} contrarily to the experimental results [see Eq.\ (\ref{expresult2moments})]. A dependence with $\gamma$ is obtained when using an colored type of noise for $\zeta(t)$, such as the Orstein-Ulhembeck (OU) one of Eq.\ (\ref{model2}) \cite{VanKampen}. The colored noise indeed introduces a typical frequency needed to simulate the frequency cut-off $\lambda$ experienced by the low-passed filtered Gaussian white noise in the experiment.

\begin{figure}[ht]
\centerline{
  \epsfysize=70mm
  \epsffile{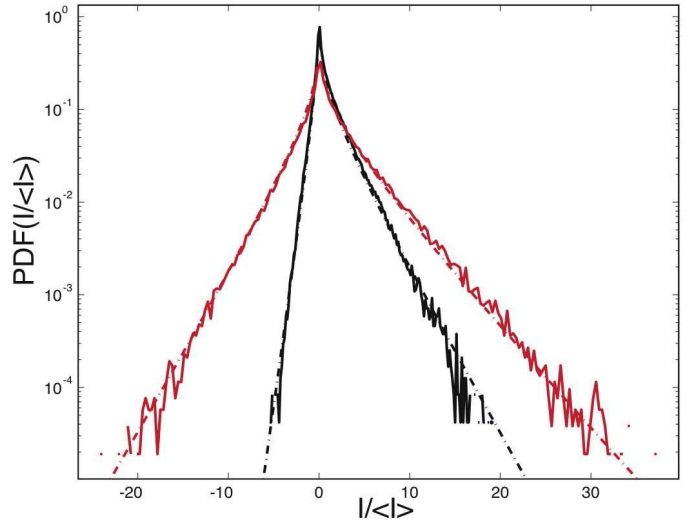}
}
 \caption{(Color online) PDFs of $I/\left< I \right>$: Comparison between experiment  ($-$) and theory [($- \cdot -$) from Eq.\ (\ref{FFApdf})] for two different values of the damping rate $\gamma=2000$ Hz ($r=\frac{< I >}{\sigma_{V}\sigma_{\zeta}}=0.45$) [black line] and  $\gamma=200$ Hz ($r=0.15$) [red (light gray) line]. The cut-off frequency $\lambda$ is fixed to 10 kHz.}
 \label{Fig7c}
 \end{figure}

As shown in Sect.\ \ref{sec:setup}, the injected power in the circuit writes $I(t)=\gamma \zeta(t) V(t)$. Using the fact that both variables $V(t)$ and $\zeta(t)$ are Gaussian with zero mean, the PDF($I$) can be written in an explicit way \cite{WT}. Let us rapidly recall the main points of its derivation. First, the stationary joint PDF($V$,$\zeta$) writes as a Gaussian bivariate  which depends only on the correlation coefficient $r \equiv \left< \zeta V \right>/\sigma_{V}\sigma_{\zeta}$ between both random variables \cite{Risken}, where $\sigma_{\zeta}=\sqrt{D/(2\lambda)}$ and $\sigma_{V}$ are the rms values of $\zeta(t)$ and $v(t)$, respectively. Second, by means of a change of variables, the PDF($\tilde{I}\equiv\zeta V=I/\gamma$) then is computed as \cite{WT}
\begin{equation}
P(\tilde{I})=\frac{\sqrt{1-r^2}}{\pi c}\exp{\left[\frac{r\tilde{I}}{c} \right]}K_0\left[\frac{| \tilde{I} |}{c} \right]
\label{FFApdf}
\end{equation}
where $c=(1-r^2)\sigma_V\sigma_\zeta$, and $K_0[\cdot]$ is the zeroth order modified Bessel function of the second kind. One have also $r=\sqrt{\gamma/(\gamma+\lambda)}$ \cite{WT}, meaning that, at fixed $\lambda$, $r$ is directly related to the damping coefficient $\gamma$. Eq.\ (\ref{FFApdf}) then is determined once $r$ is known, i.e. when $\left< I \right>$, $\sigma_V$ and $\sigma_\zeta$ are known.  Since these quantities are measured, we can compare the theoretical PDF($I$) of Eq.\ (\ref{FFApdf}) with the experimental one with no adjustable parameter. This is shown in Fig.\ \ref{Fig7c} for two different values of $\gamma$ (or $r$). The computed PDFs display a cusp at $I=0$ and exponential asymmetrical tails for large values of $I$ in good agreement with the experimental shapes. As shown in Fig.\ \ref{Fig7c}, increasing the damping rate $\gamma$ leads to PDF more and more asymmetrical with less and less negative events. The asymmetry then increases when the damping rate $\gamma$ increases. The asymmetry or the skewness of the injected power distribution is then controlled by the damping parameter $\gamma$ (or the correlation coefficient $r$ at fixed cut-off frequency $\lambda$).

For other dissipative out-of-equilibrium systems showing energy flux fluctuations, an analogue of the parameter $\gamma$ can be found. For instance, in an experiment of wave turbulence on a fluid surface \cite{WT}, the distribution shapes of the injected power $I$ by the wavemaker resemble to the ones found here: When the fluid used  is mercury, the PDF($I$) is strongly asymmetrical whereas with water, it is much more symmetrical. This is due to mean dissipation which is different for each fluid. The analogue of the $\gamma$ parameter for wave turbulence experiment is indeed related to the inverse of a typical damping time of the wavemaker which is linear with the fluid density \cite{WT}. 

With the Langevin-type model of Eqs. (\ref{model1} -- \ref{model2}), one can also calculate the first cumulants of $I(t)$. By solving the linear part of Eqs. (\ref{model1} -- \ref{model2}), the first cumulants of $I(t)$ in the stationary limit read \cite{Risken}
       \begin{equation}
        \langle I \rangle = \gamma^2\frac{D\lambda}{\lambda+\gamma},
    \label{OUresults1}
    \end{equation}

    \begin{equation}
    \sigma_{I}= \gamma^2\frac{D\lambda}{\lambda^{1/2}\gamma^{1/2}}.
    \label{OUresults2}
     \end{equation}
In the limit $\gamma/\lambda<<1$, Eq.\ (\ref{OUresults1}) yields
     \begin{equation}
        \langle I \rangle \sim D \gamma^2 \lambda^{0},
    \label{OUresults1red}
     \end{equation}
which does not depend on the cut-off frequency $\lambda$, and Eq.\ (\ref{OUresults2}) yields
     \begin{equation}
    \sigma_{I} \sim D\gamma^{3/2}\lambda^{1/2}
    \label{OUresults2red}
     \end{equation}
The range of $\gamma$ used experimentally is between 50 and 2000 Hz, and the frequency cut-off $\lambda$ is in the range from 3 kHz to 40 kHz. This leads to $\gamma/\lambda \sim 0.1$ in the worst case. The first two cumulants of Eqs.\ (\ref{OUresults1red}) and (\ref{OUresults2red}) derived from the OU process thus are in good agreement with the experimental results of Eqs.\ (\ref{expresult2moments}).

\section{Relation with the fluctuation theorem}
\label{sec:FT}
The smoothing average of the injected power $I_{\tau}$ is computed from the previous data of $I$ as
 \begin{equation}
I_{\tau}(t)=\frac{1}{\tau}\int_{t}^{t+\tau}I(t^{\prime})dt^{\prime},
\label{Itau}
 \end{equation}
where $\tau$ stands for the time of average of the signal, which is several times the correlation time $\tau_{c}$ of the injected power $I$. For our experiment, the correlation time $\tau_{c}$ is the inverse of the cut-off frequency, $1/\lambda$, which is now fixed to $10^{-4}$ s.

To describe the asymmetry of time-averaged injected power $I_{\tau}$ distribution, the quantity $\rho(\epsilon)$ is computed as
\begin{equation}
 \rho(\epsilon)\equiv \lim_{\tau\rightarrow\infty}\frac{\tau_c}{\tau}\ln{\left[\frac{P(\epsilon)}{P(-\epsilon)}\right]},
\label{rho}
\end{equation}
where $P(\epsilon\equiv I_{\tau}/\langle I \rangle)$ is the probability to have a $\epsilon$ equal to a certain value $I_{\tau}/\langle I \rangle$.  $\rho(\epsilon)$ is usually called the asymmetrical function \cite{Farago02}. Equation\ (\ref{rho}) is called the Fluctuation Theorem which states that, for times $\tau$ larger than $\tau_{c}$, this function depends only on $\epsilon$ \cite{Evans93,Evans94,Searles00}. In a certain limit, Eq.\ (\ref{rho}) takes the form
 \begin{equation}
       \rho(\epsilon) = \beta \epsilon
\label{GCeq}
 \end{equation}
where $\beta$ is a dimensionless constant. It means that the probability ratio to have a positive value of injected power ($\epsilon$) with respect to its negative value ($-\epsilon$) increases exponentially with $\epsilon$ at large $\tau$. Note that a similar relation called the Gallavotti--Cohen relationship has been derived under specific conditions \cite{GCohen}. The hypotheses for deriving Eq.\ (\ref{GCeq}) are three: the system should be microscopically reversible, dissipative and the dynamics on the phase space should be chaotic \cite{Evans93,Evans94,Searles00}. For our dissipative system, the reversibility condition is obviously not fulfilled. However, let us try to test the relation of Eq.\ (\ref{GCeq}) with our experimental data of injected power.

\begin{figure}[h]
\centerline{
  \epsfysize=70mm
  \epsffile{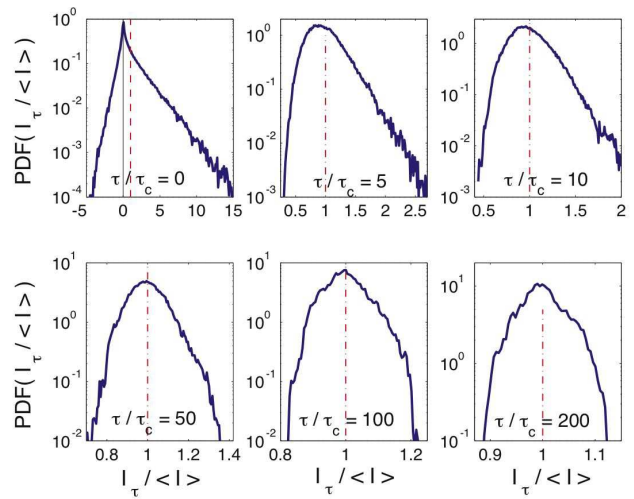}
}
  \caption{(Color online) PDF of $I_{\tau}/\langle I \rangle$ for various values of $\tau/\tau_c =$ 0, 5, 10, 50, 100 and 200 at a fixed value of $\gamma =$ 2000 Hz. The straight line ($-$) correspond to $I_{\tau}/\langle I \rangle=0$ and the dashed line ($--$) to $I_{\tau}=\langle I \rangle$.}
  \label{lissage1}
  \end{figure}

Figure \ref{lissage1} displays the PDF of time-averaged injected power $I_{\tau}/\langle I \rangle$ when $\tau/\tau_{c}$ is increased. Several features appear. First, the negative injected power events decrease with increasing $\tau$ until they disappear for $\tau\gtrsim 5\tau_c$. Second, when $\tau/\tau_{c}$ is increased, the PDF shape for negative values of $I_{\tau}/\langle I \rangle$ change from an exponential shape to a Gaussian one, whereas the exponential shape of the positive part is quite robust. Only when $\tau>>\tau_c$, the PDF shape close to the maximum tends towards a Gaussian, as one would expect from the central limit theorem.  In Fig.\ \ref{lissage1}, when $\tau/\tau_{c}$ increases, the PDF most probable value $\epsilon^*$ (i.e., where the PDF amplitude is maximum) increases slowly from $I_{\tau}/\langle I \rangle=0$ to 1 (the mean value of the injected power). This dependence of $\epsilon^*$ is shown in Fig.\ \ref{mpv} as a function of $\tau/\tau_{c}$. This dependence will be of fundamental importance when probing the FT (see below).

  \begin{figure}[h]
\centerline{
  \epsfysize=70mm
  \epsffile{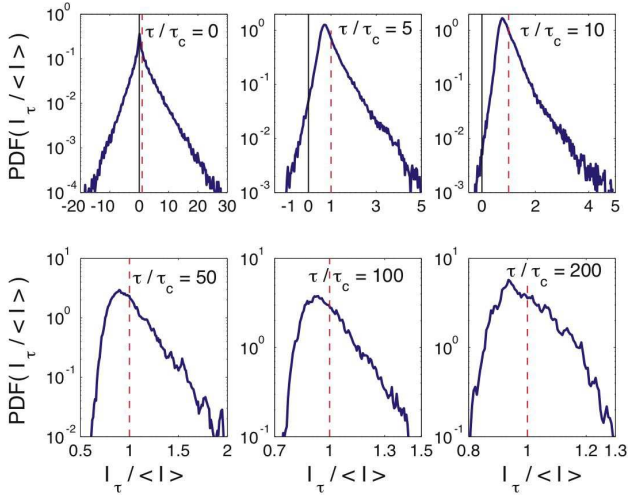}
}
  \caption{(Color online) Same as Fig.\ \ref{lissage1} for $\gamma =$ 200 Hz.}
  \label{lissage2}
  \end{figure}

The Large Deviation Function (LDF) $f(\epsilon)$ is generally defined as
 \begin{equation}
         f(\epsilon)\equiv\lim_{\tau\rightarrow\infty} \frac{\tau_c}{\tau}\ln{\left[P\left(\epsilon\equiv I_{\tau}/\langle I \rangle\right)\right]},
      \label{LDF}
   \end{equation}
and Eq.\ (\ref{rho}) thus leads to
\begin{equation}
\rho(\epsilon)=f(\epsilon) - f(-\epsilon).
\label{GCLDF}
\end{equation}
The LDF describes the probability of very large and uncommon events of $\epsilon$. It is consequently very hard to measure it. The computed LDF as in Eq.\ (\ref{LDF}) approaches its theoretical limit only for large values of $\tau/\tau_c$. With our experimental data, one can probe large values of  $\tau/\tau_c$ and therefore calculate a very accurate estimate of the LDF. Developing Eq.\ (\ref{GCLDF}) up to first order in $\epsilon$, that means, regarding only the terms $\epsilon \simeq 0$, thus leads easily to verify Eqs.\ (\ref{rho}) and\ (\ref{GCeq}). This was first conjectured by Auma\^{i}tre el al. \cite{Aumaitre99} and then predicted in a particular system by Farago \cite{Farago02}. But, what would happen if $\epsilon$ was far from zero? 

The experimental values of the asymmetrical function $\rho(\epsilon)$ are shown in Figs.\ \ref{GCexp1} and \ref{GCexp2} for two different values of $\gamma$, as a function of $\epsilon$ with $0\leq \epsilon < 3$. For small $\epsilon$, $\rho(\epsilon)$ increases linearly as expected, then $\rho(\epsilon)$ saturates when $\epsilon$ is increases further. For each value of $\tau/\tau_c$, the beginning of the saturation occurs for a critical $\epsilon$ value called $\epsilon_c <  1$. Thus, the linear prediction $\rho(\epsilon) \sim \epsilon$ is valid at finite $\tau$ only for $\epsilon < \epsilon_c$. It is important to notice that the saturation value $\epsilon_c$ of Figs.\ \ref{GCexp1} and \ref{GCexp2} corresponds to the maximum value $\epsilon^*$ of the PDF (see Fig.\ \ref{mpv}). The fact that $\rho(\epsilon) \nsim \epsilon$ for values of $\epsilon$ greater than $\epsilon_c =\epsilon^*$ is due to the different shapes of the PDF($\epsilon$) for $\epsilon < -\epsilon^*$ and for $\epsilon > \epsilon^*$ (see Figs.\ \ref{lissage1} and \ref{lissage2}). By extension to non finite $\tau$, this thus means that the FT relation of Eq.\ (\ref{GCeq}) does not hold for values of energy flux greater than its most probable value $\epsilon^*$.

\begin{figure}[t]
\centerline{
  \epsfysize=70mm
  \epsffile{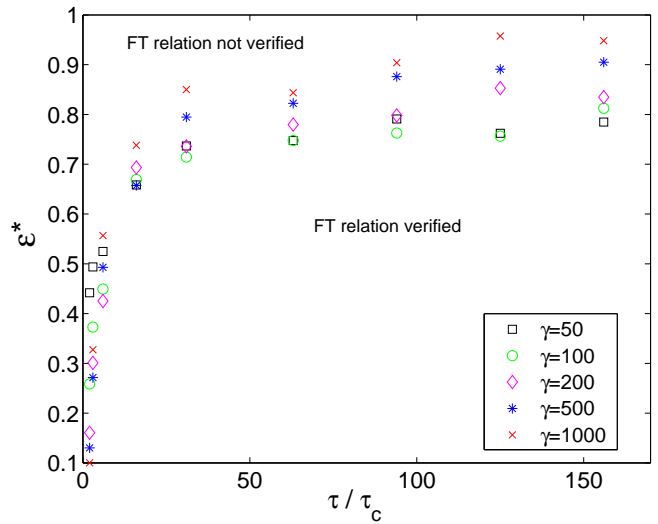}
}
  \caption{(Color online) Most probable value $\epsilon^*$ of PDF($I_{\tau}/\langle I \rangle$) as a function of $\tau/\tau_{c}$ for $\gamma =$ 50, 100, 200, 500 and 1000 Hz. For $\epsilon<\epsilon^*$, the relation $\rho(\epsilon) \sim \epsilon$ is verified at finite $\tau$, whereas it does not hold for $\epsilon>\epsilon^*$ (see text).}
  \label{mpv}
  \end{figure}

When $\tau/\tau_c$ increases, it is predicted that the most probable value $\epsilon^*$ of the PDF converges to 1 slowly as a power-law of $\tau/\tau_c$ \cite{breaking}. This power-law dependence is not found experimentally with our data (not shown here), and Fig.\ \ref{mpv} clearly shows the convergence of $\epsilon^*$ to 1 as a consequence of the convergence of the computed $f(\epsilon)$ towards the LDF. An analytical prediction for the LDF of the injected power distribution has been derived  for a Langevin equation either with a white noise \cite{Farago02} or with a colored noise (OU) forcing \cite{Jean2}. At high $\tau/\tau_c$, the shape of our experimental PDF roughly tends towards a Gaussian (see Figs.\ \ref{lissage1}-\ref{lissage2} at $\tau/\tau_c=200$) contrarily to the asymmetrical prediction of the LDF with a white or a colored noise. However, we have to be careful during this comparison due to our low statistics at very long averaged times (see the vertical range in the Figs.\ \ref{lissage1}-\ref{lissage2} at $\tau/\tau_c=200$).

Increasing now $\gamma$, at fixed $\tau/\tau_c$, leads decreasing available values of $\epsilon$ necessary to probe the Fluctuation Theorem (see Figs.\ \ref{GCexp1} and \ref{GCexp2}). It comes from the fact that when $\gamma$ is increased, the number of negative injected power events, $\epsilon<0$, decreases ($\gamma$ controls the skewness of the PDF at a given $\tau_c\sim1/\lambda$). We stress the fact that the damping rate $\gamma$, and therefore the mean dissipation, is not chosen  in this simple experiment in an \textit{ad-hoc} manner to satisfy time-reversibility. The smoothing of the signal around $\langle I \rangle$ also decreased the number of available negative events. 

  \begin{figure}[h]
\centerline{
  \epsfysize=70mm
  \epsffile{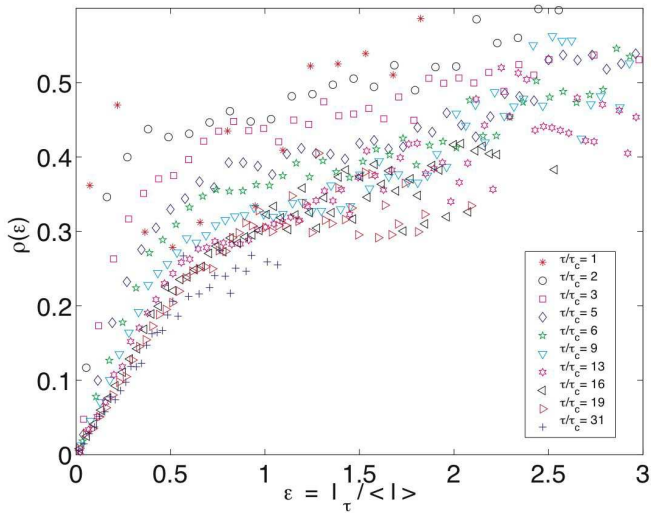}
}
  \caption{(Color online) Asymmetrical function $\rho(\epsilon) = \frac{\tau_c}{\tau}\ln{\left[\frac{P(\epsilon)}{P(-\epsilon)}\right]}$ as a function of $\epsilon$ for different integration times $\tau/\tau_c= 1$ ($\star$) to 31 ($+$) at fixed $\gamma = 100$ Hz and $D = 1.56$ mV$_{\rm rms}^{2}$/Hz.}
  \label{GCexp1}
  \end{figure}
  \begin{figure}[h]
\centerline{
  \epsfysize=70mm
  \epsffile{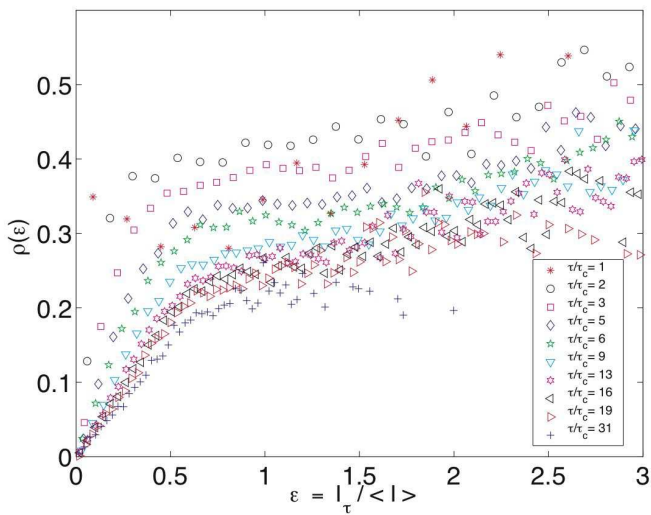}
}
  \caption{(Color online) Same as Fig.\ \ref{GCexp1} for $\gamma=50$ Hz.}
  \label{GCexp2}
  \end{figure}
In most of the previous experimental test of the Fluctuation
Theorem \cite{Feitosa04,Ciliberto98,Shang05,Ciliberto04} the limit of Eq.\ (\ref{GCeq}) is well followed, because of the small range of
explored $\epsilon \leq 0.8$ at high $\tau/\tau_c \leq 20$.
However, very recently, large range of $\epsilon$ has been measured
experimentally \cite{WT} the Fluctuation
Theorem was not satisfied. In our experiment, large range of $\epsilon$ (up to 3)
are also available even for high $\tau/\tau_c\simeq 20$. This,
thus allows us to test deeply the Fluctuation Theorem. As explained
above, the FT works only for $\epsilon$ values smaller than the
most probable value $\epsilon^*$ (see Fig.\ \ref{mpv}). Above this value saturation occurs, due to the different behavior of the PDF: for values larger than the most probable value, the PDF remains exponential, whereas for values smaller than $\epsilon^*$ it is smoother. Thus, large events of injected power are not well described by the FT, and lead to the observed saturation of $\rho(\epsilon)$.

\section{On the injected power correlations}
\label{sec:Nsyst} Dissipative stochastic systems driven out-of-equili\-brium are generally constituted of several components (e.g.,
in granular gases \cite{Feitosa04,Aumaitre99}) that may display
correlations in space and time. One can wonder how these
spatio-temporal correlations change the PDF of the injected power. Even more, it is important to study their relevance in the fulfillment of the Fluctuation Theorem.

The correlation time of the injected power into our simple experimental system can be tuned as a control parameter. To wit, the averaged injected power signal $I_{\tau}(t)$ is expressed as a sum of correlated variables where their temporal correlations mimics the spatial correlations in extended high-dimensional systems (see \S \ref{corr}). One can also look at the sum of ${\mathcal{N}}$ independent random variables distributed as $I(t)$ (see \S \ref{uncorr}). These two kind of signal processing are performed to understand if a set of statistically dependent or independent components has an effect over the fulfillment of the FT (see \S \ref{corrres}).

\subsection{Correlated components}
\label{corr}
For a single electronic circuit, the smoothing average $I_{\tau}$ of the numerically sampled injected power $I(t)$,
defined in Eq.(\ref{Itau}) can be written as the discrete sum over
$N$ points,
\begin{equation}
    I_{\tau}(t)=\frac{1}{N} \sum_{k=1}^{N} I(t+k\Delta t),
    \label{time}
\end{equation}
with $\tau\equiv N\Delta t$ and $\Delta t$ the inverse of the
sampling frequency. In our experiment, $\Delta t$ is fixed at 10
$\mu$s. Since the correlation time of the injected power,
$\tau_c\simeq1/\lambda=100\ \mu$s, is greater than $\Delta t$, the
elements of the sum above have a nonzero temporal correlation. 

 This smoothing average can be also viewed as a sum of
N statistically dependent components as
\begin{equation}
I_{\tau}(t)=\frac{1}{N} \sum_{k=1}^{N} I(t+k\Delta t)\equiv \frac{1}{N} \sum_{k=1}^{N} I_{k}(t),
\label{Correl}
\end{equation}
where $I_k(t)$ corresponds to the injected power of the $k$th correlated component.

\subsection{Uncorrelated  components}
\label{uncorr}
Let us now focus on the case where correlations between components
are neglected. That is to say each component losses its memory 
of the effect of the rest of the system faster than its internal dynamics, 
such as the case of a dilute granular gas where every particle dissipates its energy by collisions. After each collision, due to the low density of the gas, the particle losses its memory of its initial conditions
decorrelating the injected power events in time. 

We study ${\mathcal{N}}$ statistically independent variables each distributed as $I(t)$. For each time $t$, each injected power $I_i(t)$ of the $i$th non-correlated component is summed to obtain the ensemble average of the injected power,
$I_{\mathcal{N}}(t)$ defined as
\begin{equation}
     I_{\mathcal{N}}(t)=\frac{1}{\mathcal{N}} \sum_{i=1}^{\mathcal{N}} I_{i}(t) {\rm \ ,}
     \label{Uncorrel}
\end{equation}
with $I_{i}(t)$ distributed as Eq.\ (\ref{FFApdf}). This ensemble average $I_{\mathcal{N}}(t)$ should have different statistical properties than the smoothing one $I_{\tau}(t)$. Indeed, $I_{\mathcal{N}}(t)$ results from the sum over ${\mathcal{N}}$ statistically independent components [see Eq.\ (\ref{Uncorrel})], whereas $I_{\tau}(t)$ comes from the sum over N statistically dependent or correlated components [see Eq.\ (\ref{Correl})].

\begin{figure}[htb]
\centerline{
\begin{tabular}{c}
  \epsfysize=70mm
  \epsffile{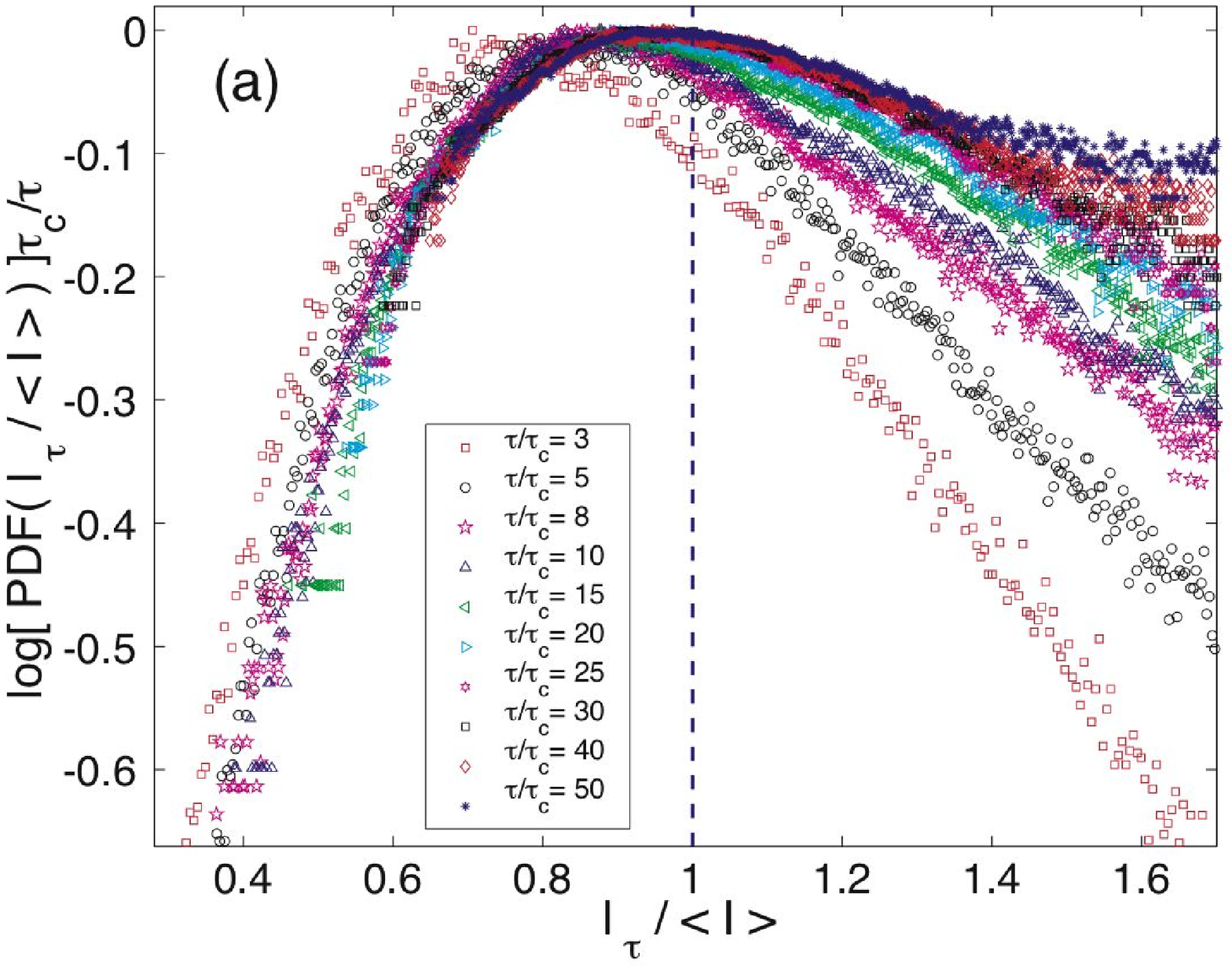}\\
  \epsfysize=70mm
    \epsffile{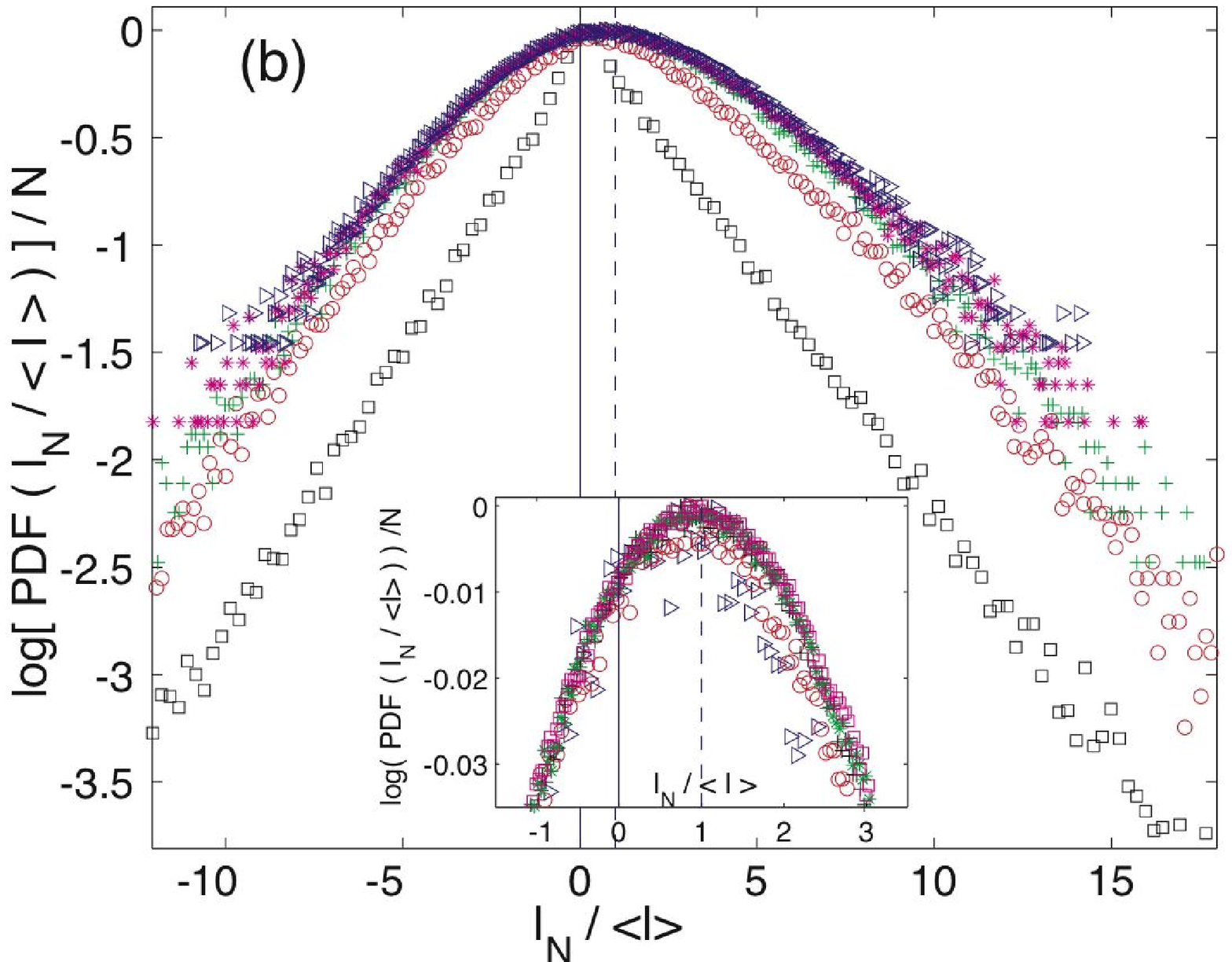}
\end{tabular}
}
  \caption{(Color online) {\bf (a)} Computed large deviation functions $\frac{\tau_c}{\tau}\ln{\left[P(I_{\tau}/\langle I \rangle)\right]}$ of $\tau/\tau_c$ correlated variables with $\tau/\tau_c=3$ ($\square$) to 50 ($\star$) for $\gamma=100$ Hz. {\bf (b)} Computed large deviation functions $\frac{1}{N}\ln{\left[P(I_{\mathcal{N}}/\langle I \rangle)\right]}$ of ${\mathcal{N}}$ uncorrelated variables with ${\mathcal{N}}=2$ ($\square$), 4 ($\circ$), 6 ($+$), 8($\ast$) and 10 ($\triangleright$) for $\gamma=100$ Hz. Inset: same with ${\mathcal{N}}=10$ ($\triangleright$), 20 ($\circ$), 30 ($+$), 40 ($\ast$) and 50 ($\square$). The dashed lines show the mean injected power $\langle I \rangle$.}
  \label{FGDNT}
  \end{figure}

\subsection{Results}
\label{corrres}
The statistical properties of the injected power into both systems described above display striking differences. Figures\ \ref{FGDNT} show the computed LDFs of $I_{\tau}$ and $I_{\mathcal{N}}$ respectively defined by $\frac{\tau_c}{\tau}\ln{\left[P(I_{\tau}/\langle I
\rangle)\right]}$ (see Eq.\ (\ref{LDF})) and by $\frac{1}{\mathcal{N}}\ln{\left[P(I_{\mathcal{N}}/\langle I \rangle)\right]}$. These LDFs describe how the fluctuations of both averages with respect to the mean $\langle I \rangle$ behave when the number of variables taken into account in the each sum, ${\mathcal{N}}$ or $\tau/\tau_c$, becomes larger and larger. The computed LDF of $I_{\tau}(t)$ is always asymmetric with exponential tails whatever $\tau >> \tau_c$, whereas the computed LDF of $I_{\mathcal{N}}(t)$ tends towards a parabola when ${\mathcal{N}}$ increases. 

For the system of statistically dependent components, the LDF of the injected power is not parabolic (as it should be if its PDF was a
Gaussian) as shown in Fig.\ \ref{FGDNT}a. The convergence to its
asymptotic shape is slow, depending strongly on the number of
components of the system (i.e., of the durations of the time
averaging, $\tau/\tau_c$). Moreover, when $\tau/\tau_c$ increases,
Fig.\ \ref{FGDNT}a shows also that the PDF's maximum slowly tends towards the mean value $\langle I \rangle$, as already noticed in the previous section (Fig.\ \ref{mpv}). As it has been already shown in Fig.\
\ref{GCexp1} in this case the Fluctuation Theorem is not satisfied.

For the ${\mathcal{N}}$ uncorrelated or statistically independent
systems, the computed LDF of $I_{\mathcal{N}}$ shown in Fig.\ \ref{FGDNT}b has exponential tails whatever the value of ${\mathcal{N}}$ as expected from the distribution of $I_{i}(t)$ [see Eq.\ (\ref{FFApdf})].
When ${\mathcal{N}}$ is increased from 2 to 10, the center of the LDF
becomes more and more parabolic (its PDF becomes more and more
Gaussian) as shown in Fig.\ \ref{FGDNT}b. This is due to the central limit theorem which can be seen as a quadratic expansion of the LDF around the mean of the distribution. At higher ${\mathcal{N}}>10$, no deviation from a parabola is observed in the inset of Fig.\ \ref{FGDNT}b due to the small fluctuation values probed. Indeed, if larger fluctuations could be accessed, one should expect exponential tails in the distribution.

\begin{figure}[h]
  \centering
\centerline{
  \epsfysize=70mm
  \epsffile{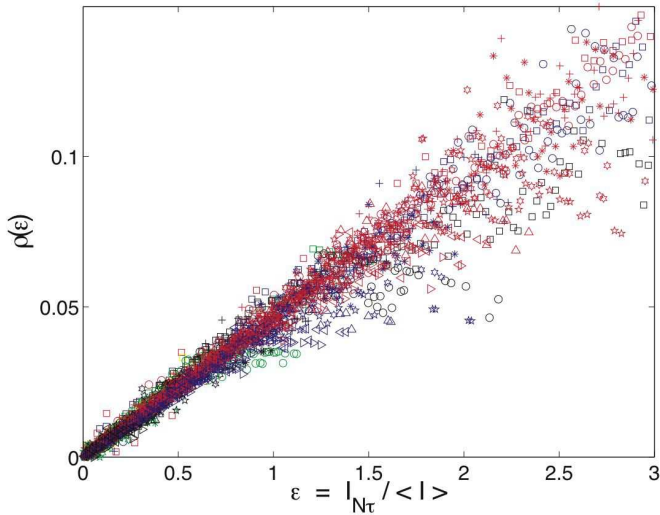}
}
  \caption{(Color online) Asymmetrical function $\rho(I_{\mathcal{N}\tau}/\langle I \rangle)$ of ${\mathcal{N}}$ independent variables, for different integration times $\tau/\tau_c=1$ ($\star$) to 31 ($+$). ${\mathcal{N}}=10$ to 100 with a 10 step.}
  \label{SansCorrelation}
  \end{figure}

Let us finally test the FT for an ensemble of ${\mathcal{N}}$
independent variables. The smoothing average of
$I_{\mathcal{N}}(t)$ over a time $\tau$ is defined as
     \begin{equation}
     I_{\mathcal{N} \tau}(t)=\frac{1}{{\mathcal{N}}\tau} \sum_{i=1}^{{\mathcal{N}}}\int_{t}^{t+\tau} I_{i}(t^{\prime})dt^{\prime}.
          \end{equation}
The asymmetrical function $\rho(I_{\mathcal{N}\tau}/\langle I
\rangle$) of the ${\mathcal{N}}$ independent variables ($10 \leq
{\mathcal{N}} \leq 100$) is plotted in Fig.\ \ref{SansCorrelation}
for 10 different integration times $\tau/\tau_c$. Whatever the
value of ${\mathcal{N}}$ and $\tau/\tau_c$,
$\rho(I_{\mathcal{N}\tau}/\langle I \rangle$) increases linearly
with $I_{\mathcal{N}\tau}/\langle I \rangle$. Thus, for the system of ${\mathcal{N}}$ uncorrelated variables ({\it e.g.} a system without spatial or temporal correlations), this means that the asymmetric function  $\rho(\epsilon=I_{\mathcal{N}\tau}/\langle I \rangle)$ satisfies the Fluctuation Theorem as soon as ${\mathcal{N}}>10$ [see Eq.\ (\ref{rho})]. However, one should be careful with this statement. Our range of accessible fluctuations $\epsilon$ decreases with increasing ${\mathcal{N}}$ and $\tau/\tau_c$. Consequently, one can only probe the Gaussian part of the PDF($\epsilon$), but not the exponential tails, leading to the linear behavior observed for $\rho(\epsilon)$. Larger $\epsilon$ values should be reached in order to observe the effect of the exponential tails on the validity of the FT.

\section{Conclusion}

    In conclusion, we have studied the statistical properties of the instantaneous injected power $I(t)$ in one of the simplest dissipative out-of-equilibrium system: an electronic $RC$ circuit submitted to a stochastic voltage. The probability distribution function (PDF) of $I(t)$ is measured for different values of the forcing amplitude and of the damping rate $\gamma$. It displays a cusp near $I\simeq 0$ and asymmetric exponential tails. This typical PDF shape has been observed in more complex dissipative systems (such as in granular gases \cite{Feitosa04}, wave turbulence \cite{WT} and convection \cite{Shang05,Shang03}). The relevant parameters of the system can be easily changed in our simple experiment. This leads to an heuristic understanding of the features of the injected power PDF by means of a Langevin-type model. The system response $V(t)$ and the forcing $\zeta(t)$ are indeed described by two Orstein-Ulhembeck random variables that follow linear coupled Langevin equations \cite{WT,Aumaitre08}. Their correlation coefficient $r=\langle V \zeta \rangle/\sigma_V\sigma_\zeta$ (related directly to the mean injected power) is the only control parameter driving the asymmetry of the distribution of $I(t)$: The larger the damping rate $\gamma$, the larger $r$, and the larger the asymmetry of the PDF. Moreover, from this model, the scaling of $\langle I \rangle$ and $\sigma_I$ are found in good agreement with the experimental measurements.
    
 The fluctuation theorem (FT) has then been probed by measuring the asymmetrical function $\rho(\epsilon)$ with $\epsilon=I_{\tau}/\langle I \rangle$, and  $I_{\tau}$ the smoothing average on a time lag $\tau$. Contrarily to previous experiments, the range of available fluctuation amplitude is large ($\epsilon \simeq 3$) even for long averaging time ($\tau/\tau_c\simeq 20$). This experiment thus allow to probe the FT in the limit of large $\epsilon$ and large $\tau/\tau_c$. We have found out that the FT is only satisfied for values of $\epsilon$ smaller than the most probable value, $\epsilon^*$ (i.e. the maximum of the PDF of $\epsilon$). For values larger than $\epsilon^*$, the asymmetrical function is no more linear with $\epsilon$ but saturates. Thus, the FT does not hold for the large available values of $\epsilon$ even at large $\tau/\tau_c$. This disagreement is not particular of this electronic system, but seems to be generic to other systems. It has been also recently observed with a wave turbulence experiment \cite{WT}. This model experiment thus appears as a useful tools to probe the FT in different limits of averaging time and fluctuation amplitudes.

Finally, this electronic experiment can be extend to mimic the behavior of a more complex out-of-equilibrium systems. To wit, we have studied the injected power fluctuations in i) a system of $N$ statistically independent components and ii) a system of ${\mathcal{N}}$ statistically dependent components. This latter can be viewed as an archetype of a dilute granular gas of uncorrelated particles. The Fluctuation Theorem (FT) for the time-averaged injected power has then been tested for the case of the correlated and uncorrelated systems. In the presence of non-zero correlation between components the FT is not satisfied, whereas it is satisfied for the uncorrelated system for finite average time $\tau$. In this last case, the fulfillment of the relation is just a consequence of the central limit theorem. Finally, this work also points out that the agreement with the FT relation is dependent on how the averaging process is performed (non overlapping bins of duration $\tau>\tau_c$ \cite{Feitosa04} or overlapping ones are two different processes related to respectively statistically independent or dependent components of the system under study).

\begin{acknowledgments}
We thank S. Fauve for suggesting us this experiment. We acknowledge S. Auma\^{i}tre, J. Farago, S. Fauve and F. P\'etr\'elis for fruitful discussions. This work has been supported by CONICYT and by ANR Turbonde No. BLAN07-3-197846.
\end{acknowledgments}

\end{document}